\newif\ifpublic
\publictrue

\documentclass{llncs}

\usepackage[utf8]{inputenc}

\usepackage[url=true,backend=bibtex8,maxbibnames=9,maxcitenames=9]{biblatex}
\usepackage{url}
\usepackage{placeins}

\usepackage[pdftex]{graphicx}

\usepackage{tikz}
\usetikzlibrary{decorations.pathreplacing}
\usetikzlibrary{shapes,arrows,patterns,trees,shadows}
\usetikzlibrary{calc}

\usepackage{listings}
\usepackage{float}
\floatstyle{ruled}
\newfloat{listing}{h}{lop}
\floatname{listing}{Listing}

\usepackage{caption}

\usepackage[leftmargin=2em]{quoting}

\usepackage[colorlinks, linkcolor=blue, urlcolor=blue, citecolor=blue]{hyperref}

\def\subheading#1{\medskip\noindent{\boldmath\textbf{#1.}}~\ignorespaces}

\begin{document}

\title{Investigating SRAM PUFs\\in large CPUs and GPUs}

\ifpublic
\author{Pol Van Aubel\inst{1}
\and
Daniel J. Bernstein\inst{2,3}
\and
Ruben Niederhagen\inst{3}}

\institute{
Radboud University\\
Digital Security Group\\
P.O. Box 9010, 6500 GL Nijmegen, The Netherlands\\
\email{radboud@polvanaubel.com}
\and
Department of Computer Science\\
University of Illinois at Chicago\\
Chicago, IL 60607--7045, USA\\
\email{djb@cr.yp.to}
\and
Department of Mathematics and Computer Science\\
Technische Universiteit Eindhoven\\
P.O. Box 513, 5600 MB Eindhoven, The Netherlands\\
\email{ruben@polycephaly.org}
}
\fi

\pdfpagewidth=210 true mm
\pdfpageheight=297 true mm
\maketitle

\begin{abstract}
Physically unclonable functions (PUFs) provide data
that can be used for cryptographic purposes:
on the one hand randomness for the initialization of random-number generators;
on the other hand individual fingerprints for unique identification of specific
hardware components.
However,
today's off-the-shelf personal computers
advertise randomness and individual fingerprints
only in the form of additional or dedicated hardware.

\medskip
This paper introduces a new set of tools
to investigate whether intrinsic PUFs can be {\it found\/}
in PC components that are not advertised as containing PUFs.
In particular,
this paper investigates
AMD64 CPU registers as potential PUF sources
in the operating-system kernel, the bootloader, and the system BIOS;
investigates the CPU cache in the early boot stages;
and investigates shared memory on Nvidia GPUs.
This investigation found non-random non-fingerprinting behavior in several components
but revealed usable PUFs in Nvidia GPUs.

\end{abstract}

\ifpublic
\catcode`@=11 \def\@thefnmark{} \@footnotetext{This work was supported
by the European Commission through the ICT
program under contract INFSO-ICT-284833 (PUFFIN),
by the Netherlands Organisation for Scientific Research (NWO) under grant 639.073.005,
by the U.S.~National Science Foundation under grant 1018836,
and by the Dutch electricity transmission system operator TenneT TSO B.V.
Permanent ID of this document: {\tt 2580a85505520618ade3cd462a3133702ae673f7}.
Date: 2015.07.29.}
\fi


\section{Introduction}
Commonly used consumer computing devices, such as
desktop computers and laptop computers, need
a multitude of cryptographic primitives, e.g.,
cryptographic operations with secret keys,
keyed hash functions, secure randomness, and, in some cases, remote
attestation and identification capabilities. 
In this paper we focus on two seemingly conflicting aspects:
The generation of \emph{random bit strings},
which requires indeterministic behavior,
and
the generation of \emph{unique identifiers},
which requires deterministic behavior.

Randomness is required for several purposes in cryptography.
For example,
random bit sequences are used to generate
secret encryption keys and nonces in cryptographic protocols
in order to make them impossible for an attacker to guess.
Many cryptographic primitives assume the presence of a secure random source;
however, most processing chips are designed to be deterministic and sources
of randomness are rare~\cite{HDWH12,LHABKW12}.

Unique identifiers 
can be used to deterministically derive an identity-based cryptographic key.
This key can be used for authentication and data protection.
For example, it would be possible to use these keys as an anti-counterfeiting
measure. Bloomberg Business reports in~\cite{BBB} that ``an `epidemic'
of bogus chips, routers, and computers costs the electronics industry
up to \$100 billion annually'', and Business Wire reports in~\cite{BW} that
``as many as one in ten IT products sold may actually be counterfeit''.
Having the ability to identify a chip as legitimate
by comparing some PUF to a database provided by the manufacturer may
help reduce this problem.
As another example, it is possible to use this key for hard disk encryption:
The hard drive,
i.e., the bootloader, operating system, and user data,
are encrypted with this secret intrinsic key and
can only be decrypted if the unique identifier is available.
The identifier thus must be protected from unauthorized access.

Currently, these features are provided by accompanying the device with
dedicated hardware:
randomness is offered, e.g., by the \verb!RDRAND! hardware random number
generator;
identification, e.g., by a Trusted Platform Module (TPM).
However,
these solutions can only be used if a dedicated TPM is available in the device
or if the CPU supports the \verb!RDRAND! instruction which only recently was
introduced with Intel's Ivy Bridge CPUs.
Furthermore, they do not help in cases where the cryptographic key should be
bound to the identity of a specific chip itself.

However, for these cryptographic functionalities
additional hardware is not necessarily required:
randomness as well as identification can be derived from individual physical
characteristics inherent to a silicon circuit by the use of physically
unclonable functions (PUFs).
PUFs can be derived from, e.g., ring oscillators~\cite{GCDD02},
signal delay variations~\cite{LLGSDD04, SS10}, flip-flops~\cite{MTV08},
latches~\cite{SHO08}, and static random-access memory
(SRAM)~\cite{GKST07,BSL13}.
While most of these require dedicated circuits, SRAM is already used for
other purposes in many general-purpose, mass-market chips.

SRAM PUFs were initially identified in FPGAs.
The PUF characteristics of SRAM are derived from the uninitialized state of
SRAM immediately after power-up.
When unpowered SRAM cells are powered up, they obtain a value of~$0$ with a
certain probability $P_0$,
or~1 with probability $P_1 = 1 - P_0$.
The individual probabilities of each SRAM cell depend on minor manufacturing
differences and are quite stable over time.
Some of the cells have a probability close to~$1$ for either $P_0$ or $P_1$
and thus tend to give the same value at every power-up.
Because of this stability, and because the pattern of this stability is
different for every block of SRAM, they can be used for fingerprinting.
Other cells have a probability close to~$0.5$ for both $P_0$ and $P_1$
and thus tend to give a different value at each power-up.
Since their behavior is unstable,
they are a good source for randomness.

Before the power-up state of SRAM can be used as PUF, an enrollment phase is
required:
the SRAM is powered up several times in order to measure which SRAM cells are
suitable for randomness and which for fingerprinting.
For the actual use of the SRAM PUF some postprocessing is performed,
e.g., a feedback loop can be used
in order to avoid bias in the generated random bit sequence
and an error correction code
in order to compensate for occasional bit errors in the fingerprint.

At TrustED 2013, researchers demonstrated in \cite{HLSKV13} that
SRAM-based PUFs exist in various brands of
popular microcontrollers, such as AVR and ARM, which are commonplace in mobile
and embedded devices. More recently \cite{SALK14} used this
to secure a  mobile platform.

We want to investigate the possible presence of PUFs in commonly used desktop
and laptop computers. For this purpose, the two most attractive targets are
the Central Processing Unit (CPU) and the Graphics Processing Unit (GPU),
since they are present in almost every desktop machine commonly in use, and
they are the chips most directly accessible by the software running on the
machine. Research into PUFs on GPUs was suggested independently by~\cite{CM13}.

The most common CPU architecture today for large computing devices,
such as laptop computers, desktop computers, and servers,
is the AMD64 architecture.
The AMD64 architecture, also known as x86-64 and x64, was introduced by AMD in
1999 as a backwards-compatible successor to the pervasive x86 architecture.
SRAM is used in abundance in the caches and registers of AMD64 CPUs.
Therefore, they may carry intrinsic PUFs.
In \cite{OGMNPV13} the authors propose an instruction-set extension to
utilize this SRAM to build a secure trusted computing environment within
the CPU.
However, research on existing PUFs in AMD64 CPUs appears non-existent.
The obvious question is whether such PUF capabilities are currently
also exhibited by
(i.e., available and accessible in) x86 and AMD64 CPUs.
The documentation of these processors contains a number of
statements which suggest that
--- even though such SRAM PUFs may exist ---
they are impossible to access from software running on those CPUs.

This paper introduces new tools to investigate
whether it is indeed impossible to use registers and
caches of AMD64 CPUs as PUFs. The result of our investigation is a negative
one, in the sense that for the specific CPU we investigated fully (an AMD E350)
we have to confirm that even at the earliest boot stages we cannot use
registers or caches as PUFs.

However, the situation is vastly different for older-generation GPUs.
Many desktop and laptop computers include hardware dedicated to processing
computer graphics, the GPU. The chips on this hardware are tailored toward
parallel computation for graphics processes (e.g., vectorized floating-point
operations), rather than the general-purpose computation done in CPUs.
Typically, GPUs have large amounts of SRAM. Contrary to the CPU, which
provides security features such as memory protection and therefore has
clear reasons to prevent direct access to the SRAM, GPUs often expose their
SRAM directly to the programmer, and also do not have the same reasons
to clear the SRAM after reset.
GPU memory and registers leak sensitive data between processes,
as observed in~\cite{Sch14} and later in~\cite{DLV13};
the absence of memory zeroing between processes, where sensitive data may
be handled, suggests that zeroing to prevent reading \emph{uninitialized}
memory is also absent.

We therefore think that it will be easier to
find and read uninitialized SRAM on GPUs than on CPUs. In this
paper we explore the possibilities for this on the Nvidia GTX 295 and find
that it is indeed possible to extract enough uninitialized SRAM to build
PUFs.
On the other hand,
we did not find PUFs on a newer generation GPU.

To enable reproducibility of our results, and to allow other researchers to
investigate other CPUs, we place all our modifications to the software
described in this paper into the public domain.
The source code and patches 
in appendix~\ref{app:src}
are available at
\url{https://www.polvanaubel.com/research/puf/x86-64/code/}.

This paper is structured as follows:
In the next section, we describe our experimental setup for the CPU,
i.e., the AMD64
processor architecture and our test mainboard, the ASRock E350M1.
In Section~\ref{sec::registers} we describe how we investigate if CPU
registers can
be accessed sufficiently early in the boot process in order to read their
power-on state and use them as SRAM PUFs.
In Section~\ref{sec::cache} we investigate the suitability of the CPU
cache as SRAM PUF during BIOS execution when the processor is in the
\emph{cache-as-RAM} mode.
In Section~\ref{sec::gpuexpsetup} we describe the experimental setup for
the GPU, i.e., the Nvidia GTX 295 GPU architecture.
Finally, in Section~\ref{sec::gpupuf} we describe the experiments conducted
on the GPU.
Finally, in Section~\ref{sec::discussion} we discuss our results.


\section{Experimental setup for the CPU}

Our main experimental setup consisted of a single mainboard with an AMD64 CPU.

\begin{figure}[t]
  \centering
  \scalebox{0.85}{\begin{tikzpicture}[font=\sffamily\small]
  \tikzstyle{box}=[rounded corners, fill=black!65, text=white, draw];
  \tikzstyle{slot}=[fill=black!25, draw];

  \node[box, minimum width=1.5cm, minimum height=1.5cm] (cpu) at (0,0) {CPU};

  \node[box, minimum width=2.25cm, minimum height=2.25cm, anchor=north, text width=2cm, align=center]
      (nb) at ([yshift=-1.5cm]cpu.south) {Northbridge\\[3mm](Memory\\Controller)};

  \draw (cpu) -- node[anchor=west, text width=\widthof{Front-Side}, align=center] {Front-Side\\Bus} (nb);

  \node[box, minimum width=2.75cm, minimum height=2.75cm, anchor=north, text width=2.5cm, align=center]
      (sb) at ([yshift=-1.75cm]nb.south) {Southbridge\\[3mm](I/O Controller)};

  \draw (nb) -- node[anchor=east, text width=\widthof{Internal}, align=center] {Internal\\Bus} (sb);

  \draw (nb.east) -- ([xshift=2.4cm]nb.east);

  \draw (nb.east) -- node[text width=1.7cm, align=center, anchor=west, at start] 
          {Memory\\Bus} ([xshift=2.8cm]nb.east);

  \node[slot, minimum height=2cm, minimum width=2mm, anchor=west] at ([xshift=1.9cm]nb.east) {};
  \node[slot, minimum height=2cm, minimum width=2mm, anchor=west] at ([xshift=2.2cm]nb.east) {};
  \node[slot, minimum height=2cm, minimum width=2mm, anchor=west] at ([xshift=2.5cm]nb.east) {};
  \node[slot, minimum height=2cm, minimum width=2mm, anchor=west] at ([xshift=2.8cm]nb.east) {};

  \draw (nb.west) -- node[text width=2.3cm, align=center, anchor=east, at start] 
            {High-Speed Bus\\AGP or PCIe} ([xshift=-2.8cm]nb.west);
  \node[slot, minimum height=3cm, minimum width=2mm, anchor=east] at ([xshift=-2.5cm]nb.west) {};
  \node[slot, minimum height=3cm, minimum width=2mm, anchor=east] at ([xshift=-2.8cm]nb.west) {};

  \draw (sb.west) -- node[text width=1.45cm, align=center, anchor=south east, at start] 
            {PCI Bus} ([xshift=-2.55cm]sb.west);
  \node[slot, minimum height=3cm, minimum width=2mm, anchor=east] at ([xshift=-1.65cm]sb.west) {};
  \node[slot, minimum height=3cm, minimum width=2mm, anchor=east] at ([xshift=-1.95cm]sb.west) {};
  \node[slot, minimum height=3cm, minimum width=2mm, anchor=east] at ([xshift=-2.25cm]sb.west) {};
  \node[slot, minimum height=3cm, minimum width=2mm, anchor=east] at ([xshift=-2.55cm]sb.west) {};

  \coordinate (sb0) at ([yshift= 12.5mm]sb.east);
  \coordinate (sb1) at ([yshift=  7.5mm]sb.east);
  \coordinate (sb2) at ([yshift=  2.5mm]sb.east);
  \coordinate (sb3) at ([yshift=- 2.5mm]sb.east);
  \coordinate (sb4) at ([yshift=- 7.5mm]sb.east);
  \coordinate (sb5) at ([yshift=-12.5mm]sb.east);

  \node[anchor=west, box, text width=15mm, align=center, minimum height=15mm]
        (obc) at ([xshift=5mm, yshift=8.725mm]sb.north east) {Onboard\\Graphics\\Controller};

  \coordinate (obc0) at ([yshift=5mm] obc.east);
  \coordinate (obc1) at (             obc.east);
  \coordinate (obc2) at ([yshift=-5mm]obc.east);

  \draw ([xshift=-5mm]sb.north east) |- 
        node[text width=7mm, align=center, near end] {PCI\\Bus} (obc.west);

  \node[box, anchor=east, text width=1.5cm, align=center, 
        minimum height=1.5cm, minimum width=1.5cm] (bios) 
      at ([xshift=-5mm,yshift=-1.625cm]sb.south) {Flash ROM\\BIOS};

  \node[box, anchor=west, text width=1.5cm, align=center, minimum height=2.75cm, minimum width=2.75cm] 
       (sio) at ([xshift=5mm,yshift=-1.625cm]sb.south) {Super I/O};

  \draw (sb.south) |- node[at start, anchor=north east, yshift=-2mm] {LPC Bus} (bios);
  \draw (sb.south) |- (sio);

  \coordinate (sio0) at ([yshift=12.5mm]sio.east);
  \coordinate (sio1) at ([yshift= 7.5mm]sio.east);
  \coordinate (sio2) at ([yshift= 2.5mm]sio.east);
  \coordinate (sio3) at ([yshift=-2.5mm]sio.east);
  \coordinate (sio4) at ([yshift=-7.5mm]sio.east);
  \coordinate (sio5) at ([yshift=-12.5mm]sio.east);

  \coordinate (tl) at ([yshift=5mm, xshift=-45mm]cpu.north);
  \coordinate (tr) at ([yshift=5mm, xshift=45mm]cpu.north);
  \coordinate (tmp) at ( sio.south -| tr);
  \coordinate (br) at ([yshift=-5mm]tmp);
  \coordinate (bl) at (br -| tl);

  \draw (tl) -- (tr) -- (br) -- (bl) -- (tl);

  \draw (obc0) -- node[at end, anchor=west] {DVI}     (obc0 -| tr);
  \draw (obc1) -- node[at end, anchor=west] {VGA}     (obc1 -| tr);
  \draw (obc2) -- node[at end, anchor=west] {$\dots$} (obc2 -| tr);

  \draw (sb0) -- node[anchor=west, at end] {Audio}    (sb0 -| tr);
  \draw (sb1) -- node[anchor=west, at end] {Ethernet} (sb1 -| tr);
  \draw (sb2) -- node[anchor=west, at end] {IDE}      (sb2 -| tr);
  \draw (sb3) -- node[anchor=west, at end] {SATA}     (sb3 -| tr);
  \draw (sb4) -- node[anchor=west, at end] {USB}      (sb4 -| tr);
  \draw (sb5) -- node[anchor=west, at end] {$\dots$}  (sb5 -| tr);

  \draw (sio0) -- node[anchor=west, at end] {Floppy Disk}   (sio0 -| tr);
  \draw (sio1) -- node[anchor=west, at end] {Keyboard}      (sio1 -| tr);
  \draw (sio2) -- node[anchor=west, at end] {Mouse}         (sio2 -| tr);
  \draw (sio3) -- node[anchor=west, at end] {Parallel Port} (sio3 -| tr);
  \draw (sio4) -- node[anchor=west, at end] {Serial Port}   (sio4 -| tr);
  \draw (sio5) -- node[anchor=west, at end] {$\dots$}       (sio5 -| tr);

  \node [anchor=south west] at ([xshift=2.5mm, yshift=2.5mm]bl) {Motherboard};

\end{tikzpicture}%
}%
  \caption[Schematic of the AMD64 motherboard architecture.]{Schematic of
          the AMD64 motherboard architecture.}
  \label{fig:motherboard_diagram}
\end{figure}
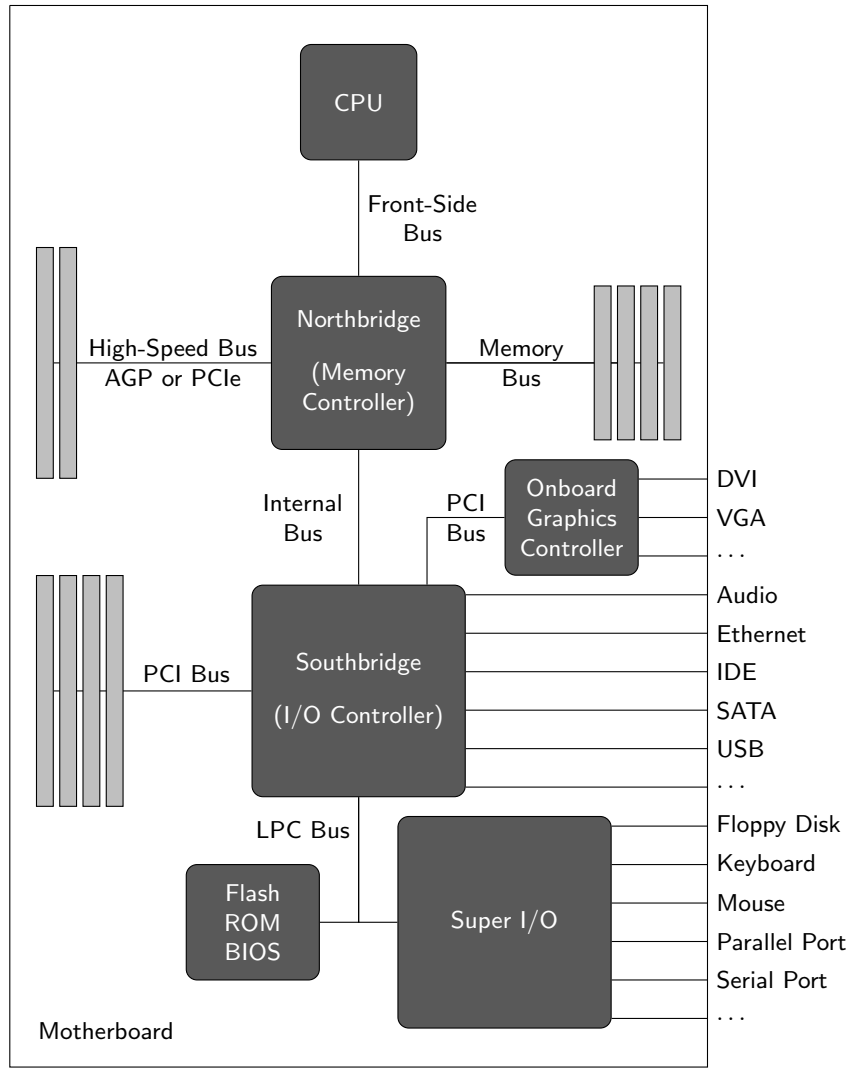

\subheading{AMD64 architecture}
Computers based on the x86 and AMD64 architectures have a long history, tracing back to
the IBM PC. The most common setup today, visualized in
Figure~\ref{fig:motherboard_diagram}, is based on a motherboard that has a socket
for an AMD64 architecture CPU, a memory controller and slots for Random Access
Memory, several communication buses such as PCI and PCI Express and associated
slots for expansion cards, non-volatile memory for storing the system's boot
firmware, and a ``chipset'' tying all these together. This chipset consists of a
Northbridge, handling communication between the CPU and high-speed peripherals
such as graphics hardware and main memory, and the Southbridge, handling
everything else, with the Northbridge as an intermediary to the CPU.

Finally, there is the Super I/O chip. This chip condenses many I/O features
which were traditionally handled by different circuits into one chip. This is
the reason that the current iteration of AMD64 motherboards still supports many
features found on boards from 20 years ago, such as serial port I/O, floppy-disk
drives, and parallel ports, next to relatively new features such as
Serial ATA and PCI Express. However, some of these features might not be exposed
to the user:
The Super I/O chip that is used to drive these subsystems often supports
the entire range of ``old'' functionalities, but only those which the
motherboard manufacturer deems worthwhile to offer are actually exposed through
sockets on the board. The serial port, for example, is still exposed as a
header on most boards, or at least as a solder-on option. Since these are
relatively simple I/O devices, they are often the first to be initialized after
system startup and can be used for output of, e.g., system diagnostics during
the early boot stage before the graphics hardware has been initialized.

In recent years, functions the Northbridge used to handle, such as
memory control and graphics-hardware control, were integrated into the CPU.
This was done to reduce overhead and speed limitations caused by having to go
through an intermediary chip. Since this lifted most of the high-speed demands
from the Northbridge, this development
has caused manufacturers to integrate the few remaining functions of the
Northbridge and the functions of the Southbridge into a single chip. The
main principles of operation of the motherboard, however, remain the same.

\subheading{Test mainboard}
\label{sec:mainboard}
Our main test board is the E350M1, manufactured by ASRock. On it runs an AMD
E-350 APU (Accelerated Processing Unit, a package embedding a CPU and graphics
controller) which was first manufactured in 2011, with an AMD A50M chipset. It
has an exposed serial port header and a socketed 4~MiB Winbond 25Q32FVAIQ NVRAM
chip for the UEFI or BIOS firmware.
The board has on-board flash capabilities for this chip.
The form factor is mini-ITX.
The E-350 APU itself has two processor cores,
with
32~KiB level-1 data cache,
32~KiB level-1 instruction cache,
and
512~KiB of level-2 cache per core.

As explained later in Section~\ref{sec:mainboardreqs}, the main reasons for
picking this mainboard are that it supports a fairly recent AMD CPU, has a
socketed NVRAM chip, and is supported by the open-source BIOS implementation
coreboot~\cite{coreboot}.

The integration of graphics hardware,
combined with the small form factor, make this a board suited for
general-purpose home computing and multimedia computers.

\begin{figure}[t]
	\center
	\includegraphics[width=0.6\textwidth]{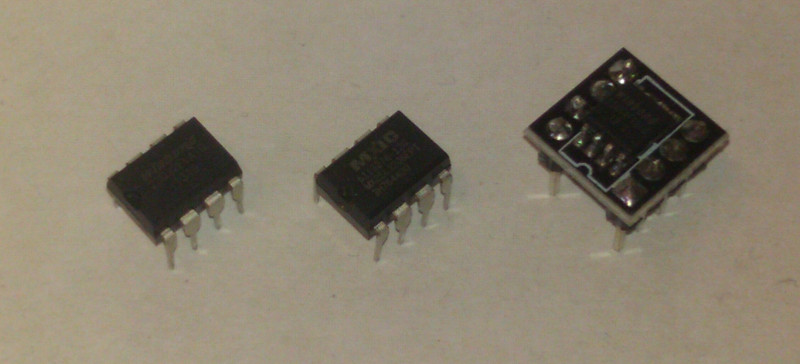}
	\caption[Chips used on the E350M1 motherboard.]{Chips used on the E350M1
		motherboard. Left: the original Winbond 25Q32FVAIQ. Center:
		The unsuitable replacement MX25L3206EPI. Right: The working
	replacement Winbond 25Q64FVSIG}
	\label{fig:chips}
\end{figure}

We acquired two sets of replacement NVRAM chips. The first set consisted of five
MXIC MX25L3206EPI. These chips closely match the original chip's specifications, yet
are from a different manufacturer. They failed to
boot the board with anything other than the original UEFI firmware. The second
set consisted of two Winbond 25Q64FVSIG chips. These chips are almost identical
to the original, with only two major differences:
they have twice the storage size (8~MiB),
and a different form factor (SOIC8 instead of DIP8).
Therefore, they required an adapter circuit to fit the form factor.
However, these chips served the purpose of booting the board with modified
firmware.
The three different types of chips can be seen in Figure~\ref{fig:chips}.
For flashing these chips under Linux, we used the open-source software flashrom.

\begin{figure}[t]
	\center
	\includegraphics[width=0.65\textwidth]{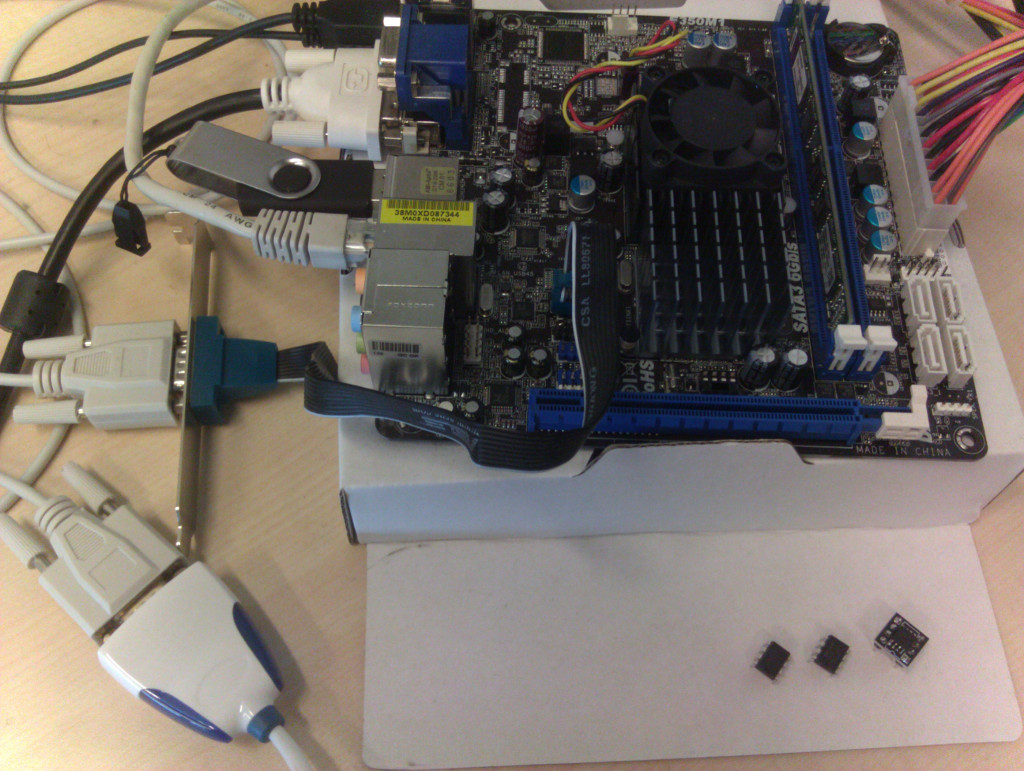}
	\caption[Photograph of the E350M1 motherboard.]{Photograph of the
		E350M1 motherboard.}
	\label{fig:e350m1photo}
\end{figure}

For mass storage (bootloader and operating system) we used a simple USB
stick. For I/O we used a normal setup of keyboard, mouse and screen, but also
attached a serial socket to the serial port header, and used a serial-to-USB
adapter to get serial output from BIOS and bootloader. The test setup can be
seen in Figure~\ref{fig:e350m1photo}.

Finally, power was supplied by a normal ATX power supply, and we powered,
unpowered and reset the board by shorting the corresponding pins with a metal
tab.
Measurements were taken by manually powercycling the board and reading the
measurement output from screen (kernel) or serial output (BIOS and bootloader).

\section{CPU registers}
\label{sec::registers}

There are indications that both Intel and AMD use SRAM to build the register
banks present in their CPUs~\cite{BOHR09},
although this is not explicitly mentioned in the specification charts for their
CPUs.
The register banks contain, among others, general-purpose registers, MMX
vector registers, and XMM vector registers.
Of these, the general-purpose registers are likely to be heavily used from the
moment of system start, since many of them are required to be used in basic
instructions.
The XMM registers, however, can only be accessed by the use of the Streaming
SIMD Extensions (SSE) instruction set, which is unlikely to be used by the
system startup code.
They are therefore good candidates to check for PUF behavior.

However, the \emph{AMD64 Architecture Programmer's Manual Volume 2: System
Programming} \cite{AMD64PM2}
contains several statements which give reason to believe that it would be 
extremely hard, if not outright impossible, to get to the power-on state of the
register banks.
For instance, Table~14-1 of that document shows the initial processor state that
follows \verb!RESET! or \verb!INIT!.
The table lists a deterministic state for all the general-purpose registers,
most of which get initialized to~0.
The 64-bit media state (MMX registers) and the SSE state
(XMM registers) are also initialized to 0 after \verb!RESET!.
After \verb!INIT!, however, they are apparently not modified, but
since it is not possible to initialize a processor without going through
power-on \verb!RESET! at the beginning, this does not help
either.
Volume~1 of the \emph{Programmer's Manual} also states that, upon power-on,
all YMM/XMM registers are cleared.
This confirms the conclusions drawn from the table in Volume~2.

Experimental results show that the register banks are indeed not usable as
PUFs on our testing machines. To explain this conclusion, we will describe
the x86/AM64 boot process, and discuss how to dump the state of the XMM
registers during different stages of the boot procedure.

\subsection{Boot process}
\label{sec:boot-process}

The boot process for an AMD64-based machine consists of several steps.
The Southbridge loads the initial firmware code (BIOS or UEFI), and the
processor starts executing from the \verb!RESET! vector (address
\verb!0xFFFFFFF0!).
This code performs CPU initialization and initialization of other mainboard
components such as the Super-IO chip, 
responsible for input-output through devices such as the serial port, and
the memory controller, responsible for driving and communicating with main memory.
Next, it searches for all bootable devices and finally loads the 
bootloader from the desired location.

The bootloader allows the user to select between different operating systems,
loads the desired operating-system kernel and any other required resources, and
then hands over control to this kernel.
From this moment on the operating system is in control.

One of the main differences between BIOS and UEFI boot options is that a BIOS
system will, in order to start the bootloader, drop the CPU back into 16-bit
real mode, whereas a UEFI system can directly load the bootloader in 32-bit
protected or 64-bit long mode.
We have looked at systems using the BIOS model,
but our findings apply to the UEFI model as well
since the UEFI model is not different from the BIOS model in how it initializes
the CPU, Super-I/O, and memory controller. 
For the rest of this paper,
when discussing bootloader and boot firmware, we assume the BIOS model.

This division of stages in the boot process is also reflected in the complexity
of the software running in each stage. The BIOS is small, very specialized, and
designed to work for specific hardware. The bootloader, in turn, is somewhat
larger, somewhat more portable, but still has a very limited set of tasks.
Finally, an operating-system kernel is often large and complex, and designed to
deal with many different hardware configurations and many different use cases.
If PUF behavior can easily be exposed at the operating system level,
without edits to the underlying layers, this enables wide deployment with
relatively little development. If, however, the BIOS needs to be edited, then
deploying a system using these PUF results would require edits to each mainboard
that the system will use. The tradeoff here is that a solution which does not
require edits to the BIOS and bootloader would implicitly trust these
components, whereas a solution where the BIOS needs to be edited would be able
to work with a much smaller trusted base system.

Because of these considerations, we decided to explore all three options. In the
following sections, we first look at the kernel level, before going to the
bootloader, and finally to the BIOS.

\subsection{Kernel}

The operating-system kernel is started by a bootloader in our test setup.
We can only be sure to read potentially uninitialized values from registers
if we read the state of the registers as early as possible,
before they are used either by the operating system
or by user processes.
Thus, the register state must be stored during the startup-process of the
operating system.
This requires us to modify the source code of the operating-system kernel.
Therefore, the obvious choice is to use an open-source kernel.
We decided to use Linux.

Our code that reads out and displays the contents of the XMM registers consists
of two parts:
a kernel patch that stores the content of the XMM registers right after those
registers have been made available
and a kernel module that gives access to the stored data
after the boot process has been finished.

\subheading{Kernel patch}
Before XMM registers can be accessed, the processor must be switched to the
correct mode using the \verb!CR0! and \verb!CR4! control
registers~\cite[Page~433]{AMD64PM2}.
This happens in \verb!fpu_init! in file
\verb!arch/x86/kernel/i387.c! of the Linux kernel.
Before this function is called, the kernel does not have access to the XMM
registers.
Thus, it is not possible that the XMM registers have been used before within the
kernel and that potential PUF data in those registers has been overwritten by
the kernel.

We are storing the data of all XMM registers into memory right after the
control registers have been set, in order to ensure that our code is the first
kernel code that accesses the registers.
We use the instruction \verb!FXSAVE! in order to save all the FPU and XMM
registers to memory at once; the kernel patch adds only 5 lines of source code.

\subheading{Kernel module}
Displaying or permanently storing data in the very early phase of the kernel
boot process is tedious.
Therefore, we simply store the data at boot time and make it available to user
space applications once the boot process is finished via a kernel module.
The kernel module provides entries (one for each CPU core) in the \verb!proc! file
system that can simply be read in order to obtain and display the XMM register
data.

\subheading{Results}
We tested our code on two AMD64-based machines,
first on a surplus office machine with an AMD Athlon~64~X2~3800.
Later, we re-ran the tests on the dedicated test-board with an AMD E350 CPU
described in Section~\ref{sec:mainboard}.
Both CPUs are dual-core CPUs.
On both boards, all XMM registers on the second CPU core contained all~0.
The registers on the first CPU core contained some data, some of it stable over
several reboots, some of it varying.
However, some of the registers obviously contained ASCII code, e.g., the strings
``\verb!GNU core!'', 
``\verb!GB.UTF-8!'', 
and ``\verb!: <%s>!''. 
This indicates that the XMM registers have been used by the boatloader --- if
not directly in the source code then maybe by C standard-library calls like
\verb!memcpy!, \verb!memcmp!, or string operations;
disassembling the GRUB boatloader 
shows many occurrences of vector instructions on XMM registers.

Thus, at the time of kernel startup, the initial status of the registers has been
modified and they cannot be used as PUF.
Therefore, in the next step we investigated the status of the XMM registers
before the kernel is started, i.e., in the early stages of the bootloader.

\subsection{GRUB}
\label{sec::grub}

The bootloader is a user-controlled piece of software, often installed into the
boot sector of one of the hard disk drives.
However, it runs still fairly early in the boot process.
This combination of factors makes it a good candidate for attempting to find
uninitialized SRAM in the XMM registers of a CPU.

\subheading{GRUB patch}
GRUB (GRand Unified Bootloader) is a free open-source bootloader for AMD64
systems~\cite{grub}. 
It is one of the most popular bootloaders used to boot Linux systems
and fairly easy to modify. 
After GRUB starts, it switches the CPU back into 32-bit protected mode as soon
as possible. Then it does some more machine initialization and checks, during
which it initializes the terminal console, either over the VGA output or serial
output. Next, it loads all the modules it requires, loads its configuration, and
displays the boot menu for the user to select an operating system.

In the previous section, we mentioned that disassembly of GRUB shows many uses
of the XMM registers.
However, at the moment when GRUB starts, the CPU is still in 16-bit real mode.
Therefore no XMM registers are available to be used.
In order to be early enough to read uninitialized registers, we changed the GRUB
source code so that immediately after machine and terminal initialization, we
enable access to the XMM registers ourselves, then
read the register contents of the XMM registers \verb!XMM0! to \verb!XMM7!.
Next, we write them to the terminal.
First we allocate a block of memory with a size of 1024 bits (128 bits for each
register) and fill it with a known pattern.
Next, we enable SSE-instructions on the CPU in the first \verb!asm!-block.
Immediately after that we copy the contents of each register to the memory
region allocated before,
in the second \verb!asm!-block. We do not use the \verb!FXSAVE! instructions
here, rather, we perform a single \verb!MOVUPD! instruction for each register we
want to store. Finally, we write the values from memory to the
console. Disassembly of the resulting GRUB image shows that, indeed, our
reading of the XMM registers is the first use of these registers within GRUB.

\subheading{Results}
Again, we tested our code on the surplus office machine described above and
later also on the dedicated test mainboard.
Unfortunately, on the first test-machine the contents of all registers except
for \verb!XMM0! were~0.
\verb!XMM0! was filled with a static value which turned out to be a fill-pattern
used in the initialization code of main memory in AMD-supplied BIOS code.
These values were stable over repeated tests.
This indicates that at this point the registers have been zeroed
and that at least register \verb!XMM0! has been used already by the BIOS.
For the same reasons as before, this means that at this point the XMM registers
cannot be used as PUF, neither for randomness nor for fingerprinting.
Therefore, as the next step we turned to the BIOS in the attempt to read data
usable as a PUF from the registers.

\subsection{Coreboot}

As stated before, the BIOS is the first code run by the CPU. 
It detects and initializes the hardware and firmware, 
puts the CPU in the correct mode, runs software that makes it possible to configure the BIOS itself,
and loads and runs the bootloader.
The BIOS is the earliest step in the boot process that can be controlled, unless one
has access to the CPU microcode. 

The BIOS is loaded from an NVRAM chip.
Often, its machine code is readable by reading out the NVRAM chip or by dumping
the contents of BIOS updates.
However, it is not easy to edit the BIOS code without access to its source code,
which most mainboard vendors do not provide.
Luckily, it is not necessary to reverse-engineer the closed-source BIOS provided
by the mainboard vendors; there is an alternative:
coreboot, formerly linuxBIOS, is a free open-source machine-initialization
system \cite{coreboot}.
It is modularly built so that it can function as a BIOS, a UEFI system, or in
several other possible configurations.

\subheading{Mainboard selection}
\label{sec:mainboardreqs}
Coreboot, despite its modularity, needs to be ported to every individual
new mainboard for which support is desired.
This is caused by subtle differences in hardware configuration, and is even
required if a board uses chips which are all already supported by coreboot.
Instead of porting coreboot to the AMD Athlon 64 X2 3800 mainboard mentioned
before that we already had ``in stock'',
we decided to acquire a board that coreboot had already been ported to by the
community;
our first requirement for the board was that it must support modern AMD64 CPUs.

Since the BIOS resides in an NVRAM chip on the mainboard, the only way to
install a new BIOS is by flashing this chip.
Most modern mainboards have this flash-capability built into the mainboard
itself and software running in the operating system can flash the BIOS in order
to enable user-friendly BIOS updates.
However, should a modification to the BIOS source code render the system
unbootable, this on-board capability will obviously not be available.
Therefore an additional requirement was that the mainboard that we were going to
use must have a socketed NVRAM chip rather than one soldered onto the board.
This would allow us to boot the board with a ``good'' chip, then switching the
chips and re-flashing the bad one.

Because of these requirements, our choice was the ASRock E350M1 mainboard described in
Section~\ref{sec:mainboard}.

\subheading{Coreboot patch}
The coreboot boot process begins the same as described in
Section~\ref{sec:boot-process}:
the Southbridge loads the coreboot image,
then the CPU starts processing from the \verb!RESET! vector.
The first thing coreboot does is to put the CPU into 32-bit protected mode.
It then does some additional CPU initialization,
initializes the level-2 cache as RAM for stack-based computing,
initializes the Super-IO chip for serial port output, and then starts outputting
diagnostic and boot progress information over the serial port.
It initializes the memory controller, and eventually it loads the payloads
stored in NVRAM,
which can vary: a VGA ROM to enable VGA output, a BIOS or UEFI implementation,
an operating-system kernel directly, or several other possibilities.

As soon as the cache-as-RAM initialization is done, memory is available to store
the values of the XMM registers.
We changed coreboot similar to how we changed GRUB.
First, we allocate a buffer of 1024 bits of memory and fill them with a known
pattern.
Then we copy the contents of the XMM registers to the buffer.
At this point, there is no interface initialized to send data out of the CPU,
except for a very rudimentary POST code interface which can send one byte at a
time and requires a special PCI card to read it. This is inconvenient at best,
so we allow coreboot to continue machine initialization until the serial
port is enabled.
Then, we write the values previously read from the registers out over the serial
console.

\subheading{Results}
This time, all the registers contain 0 on our test machine.
Manual analysis of a disassembly of the coreboot firmware image flashed to
the device shows that \verb!XMM0! and \verb!XMM1! are
at some earlier point used to temporarily store data,
but \verb!XMM2!--\verb!XMM7! are not used
before being copied by the modified code.
This matches the documentation, and implies that there is no way to get
access to uninitialized SRAM state by using XMM registers.


\section{CPU cache}
\label{sec::cache}

The AMD64 architecture defines the possibility of several levels of cache,
while leaving the exact implementation to manufacturers of actual CPUs.
As mentioned before, caches are usually implemented as SRAM.
Therefore, reading the bootup-state of cache could be another source of PUF
behavior.

\subsection{Cache operation}

During normal operation of an AMD64-based machine, main memory is available
through a memory controller.
The use of caches speeds up memory accesses by granting the CPU fast read
and write access to recently touched data which would otherwise have to be
fetched from main memory. On the AMD64 architecture, the data stored in caches
is always the result of a read from
main memory or a write to main memory; caches act as a fast temporary buffer.
It is not possible for software to explicitly write to, or read from, cache.
If software needs to use data from a certain address in main memory,
the corresponding cache line is first loaded into cache,
then accessed and potentially modified by the software,
and eventually modifications may be written back to main memory.
Thus, the cache contains a copy of the data that should be in main memory, but
that might not be the exact same data as what \emph{is} in main memory because
the writeback has not happened yet.
When exactly reads from and writes to main memory are performed,
depends on the \emph{memory type} assigned to the section of main memory being
handled.
For the purposes of this paper, we will only examine the memory type
\emph{writeback}~\cite[Page 173]{AMD64PM2}.

On multicore systems and cache-coherent multi-socket systems, another problem is
that the data in cache itself might not be the most up-to-date copy of the data.
Because of this, the cache
controller must keep track of which data is stored in which location
(a specific cache or in main memory) at what time.
In order to keep track of this, the MOESI protocol is used that allows
cache lines to be in one of five different states: 
\emph{Modified}, \emph{Owned}, \emph{Exclusive}, \emph{Shared}, and
\emph{Invalid}~\cite[Pages 169--176]{AMD64PM2}.

Many modern AMD64 CPUs support what is known as cache-as-RAM operation. This
uses the level-2 cache in each CPU core to
enable stack-based computing during the early boot process. At this point
the memory controller has not yet been initialized, so main memory is
unavailable~\cite[Pages 32--33]{BKDGE350}.
In cache-as-RAM operation mode, the memory state \emph{writeback} is assigned to
all available memory addresses.
After the CPU received a \verb!RESET! signal, the entire cache is in the state
\emph{Invalid}.
In writeback mode Invalid state, any memory read will trigger a ``read miss'',
which would normally cause a read from memory into cache, and put the cache
line in either \emph{Shared} or \emph{Exclusive} state.
Any memory write will cause a ``write miss'', since the line needs to be
modified and held as Modified in cache. Therefore, a write miss
would normally cause a read from memory,
modify the corresponding data, and put the cache line in
\emph{Modified} state~\cite[Pages 169--171]{AMD64PM2}.
However, the documentation does not state what happens when these misses are
encountered during the early boot process when the memory controller is still
disabled.
It could be the case that any
read from main memory will be handled within the CPU to return some static
value, e.g., zero. It could also be the case that the cache is not actually
modified on a read, in which case reading a block of memory might give us
the power-on state of the SRAM cells in the cache.

\subsection{Coreboot}

The cache-as-RAM initialization code used by coreboot, written by AMD, contains
instructions to explicitly zero out the cache area used as stack.
Furthermore, a comment on lines 51--58 of \verb!src/cpu/x86/16bit/entry16.inc!
(one of the source files used to define the earliest stages of the coreboot boot
process before the CPU is switched to 32-bit protected mode)
implies that coreboot used to explicitly invalidate the cache at that point, but
no longer does for performance reasons.
This could imply that power-on values from the cache are indeed readable after
cache-as-RAM initialization, if the instructions to explicitly zero the cache
are removed.

\subheading{Coreboot patch}
To test this, we replaced the instructions zeroing out the cache with
instructions filling it with a known pattern.
Then we allowed the boot process to continue until initialization of the serial
console.
As soon as the serial
console was available, we output the entire contents of the memory region
used as stack, and confirmed that the known pattern was there. This ensures
that we were modifying the correct code, and that the values were not being
changed between the initialization of the cache and the output. After this
test, we simply removed the instructions writing the pattern entirely to get
the power-on state of the SRAM. These patches to coreboot should be applied
separately from the earlier, register-related patches.

\subheading{Results}
Unfortunately, as in the previous experiments,
the output consisted mostly of zeroes, and
the parts that were non-zero were clearly deterministic and at the top of the
memory region.
This part of the memory most likely is the region of the stack that already has
been used by function calls before and during serial console
initialization.
Therefore, also cache-as-RAM does not provide access to SRAM in bootup state;
the CPU transparently takes care of wiping the cache before the first read
access.


\section{GPU experimental setup}
\label{sec::gpuexpsetup}
Our experimental setup for the GPUs consisted of several modern desktop
machines, each running one or two GPU cards based on the Nvidia GTX 295.
We used the CUDA SDK version 4.0.

\subheading{Graphics Processing}
Graphics cards used to provide only operations for graphics processing.
However, in the past decade, a shift has taken place tailored to expose
this power, providing a more general-purpose instruction set along with
heavily vectorized, parallel computation. Because of this, non-graphical
programs have started to utilize this power by offloading certain computations
to the GPU that would previously have been done by the CPU.

Graphics programming is usually done using various high-level graphics
APIs, such as OpenGL and DirectX. However, the more general-purpose use of
their operations is done through other
semi-portable high-level programming interfaces, such as CUDA~\cite{GPUCUDA}
and OpenCL.
The CPU, and therefore any normal user program, does not have direct access
to the GPU's SRAM memory. Furthermore, the public documentation for the
actual low-level instruction sets is not as extensive as for CPUs.
For example, one of the ways Nvidia card programming is done is by writing
programs in CUDA, which then compiles into still semi-portable,
high-level, ``assembly-language-like'' PTX~\cite{GPUPTX},
still hiding most of the
hardware details. The PTX is in turn compiled by the
GPU card's driver to a binary ``kernel'' which is run on the card itself.

On the other hand, GPUs evolved as single-user devices, dedicated to
processing (non-sensitive) graphics data, without many of the
security features of CPUs. Considering those features, such as virtual memory,
address space separation, and memory protection, it is unsurprising that
the CPU indeed clears its SRAM and makes it unavailable to any outside
applications. Since GPUs do not have to take this into consideration,
it is possible that there will be no logic to clear the SRAM or make
it unavailable to outside applications. On top of that, in contrast with
their instruction sets, GPU hardware tends to be documented as well as or
better than CPUs. There also exists research into the non-documented
aspects of the architecture, see e.g.~\cite{WPSM10}.

\subheading{Nvidia GTX 295 GPU card}
The Nvidia GTX 295 GPU card contains two graphics processing devices.
Each of these devices has 896MiB of DDR3 RAM --- ``global memory'' --- and
30 multiprocessors (MPs). Each of the MPs, in turn, has 8 arithmetic logic
units (ALUs), 16384 32-bit registers, and 16KiB SRAM --- ``shared memory''.
Nvidia GPUs can be programmed for general-purpose computing using the CUDA
framework.


\section{GPU multiprocessor shared memory}
\label{sec::gpupuf}

Even though more SRAM is available in the registers, the shared memory SRAM
is easier to access. The main cause of this is that CUDA and PTX make it
easy to access the shared memory through a linear address space, but
there is no real assembly language provided by NVIDIA that would allow to
directly access registers.

Using Nvidia's CUDA language, we developed an SRAM readout tool.
CUDA hides most of the hardware details, but it provides enough control to
access specified locations in SRAM.
The tool works by copying the shared memory SRAM to global memory DRAM, after
which the code running on the host CPU reads this data.
The actual size of the SRAM is 16384 bytes, but the first 24 bytes are
reserved for kernel parameters (e.g., the thread id) and for the function
parameters passed to the kernel.
Thus, only the latter $16384 - 24$ bytes can be accessed from CUDA code.
The resulting loop doing this is very simple:
\begin{lstlisting}[language=C]
     #define MAX (16384 - 24)

     __global__ void read(unsigned char *data)
     {
         __shared__ unsigned char d[MAX];

         for (int i = 0; i < MAX; i++) {
             data[blockIdx.x * MAX + i] = d[i];
         }
     }
\end{lstlisting}

\subheading{Results}
The power-on SRAM contents appear to contain large amounts of random data.
Powering off and on again produces a similar, but not identical, SRAM state.
Overwriting the SRAM state and resetting the GPU again produces a similar
state, as if the SRAM state had never been overwritten.
A different GTX 295 GPU has a different power-on SRAM state.
These observations were consistent with what one would expect from
uninitialized SRAM.

In the end, we were able to read out 490800 bytes out of the 491520 bytes
of shared memory in each GPU. We repeated this experiment on 17 devices.

\begin{figure}[h]
	\center
	\includegraphics[width=0.65\textwidth]{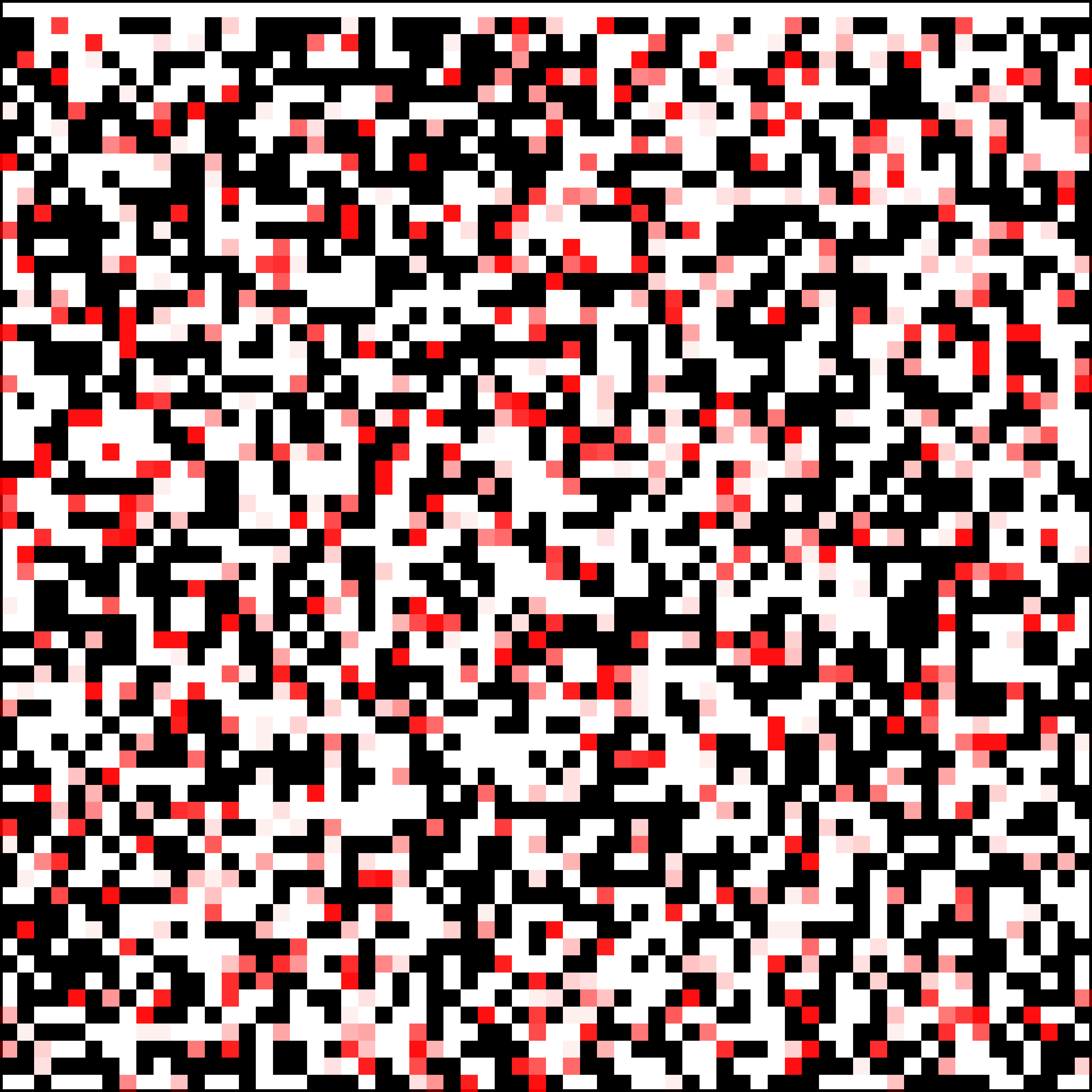}
	\ifpublic
	\caption[antilles0, device 0, MP 0, 17 traces.]{antilles0, device 0, MP 0, 17 traces.}
	\fi
	\label{fig:gpu-traces}
\end{figure}

Figure~\ref{fig:gpu-traces} shows an example of a GPU SRAM PUF from device~0,
MP~0 on
\ifpublic
the machine ``antilles0''.
\fi
We took 17 measurements, each after a power-off reboot.
The figure shows different colors for each bit of the first $64 \times 64$
bits of the SRAM; white pixels indicate that a bit was $1$ on each power-up,
black pixels indicate that the bit was $0$; different shades of red indicate
the ratio of $1$ versus $0$ on each power-on.
Thus, the corresponding bits of black/white pixels can be used to identify the
SRAM and thus the device, while the bits of the red pixels can be used to
derive randomness from the device.
The first accessible 64 bits are allways $0$ and thus appear to be cleared
on kernel launch when kernel parameters are copied to the SRAM.

\begin{figure}[h]
	\center
	\includegraphics[width=0.85\textwidth]{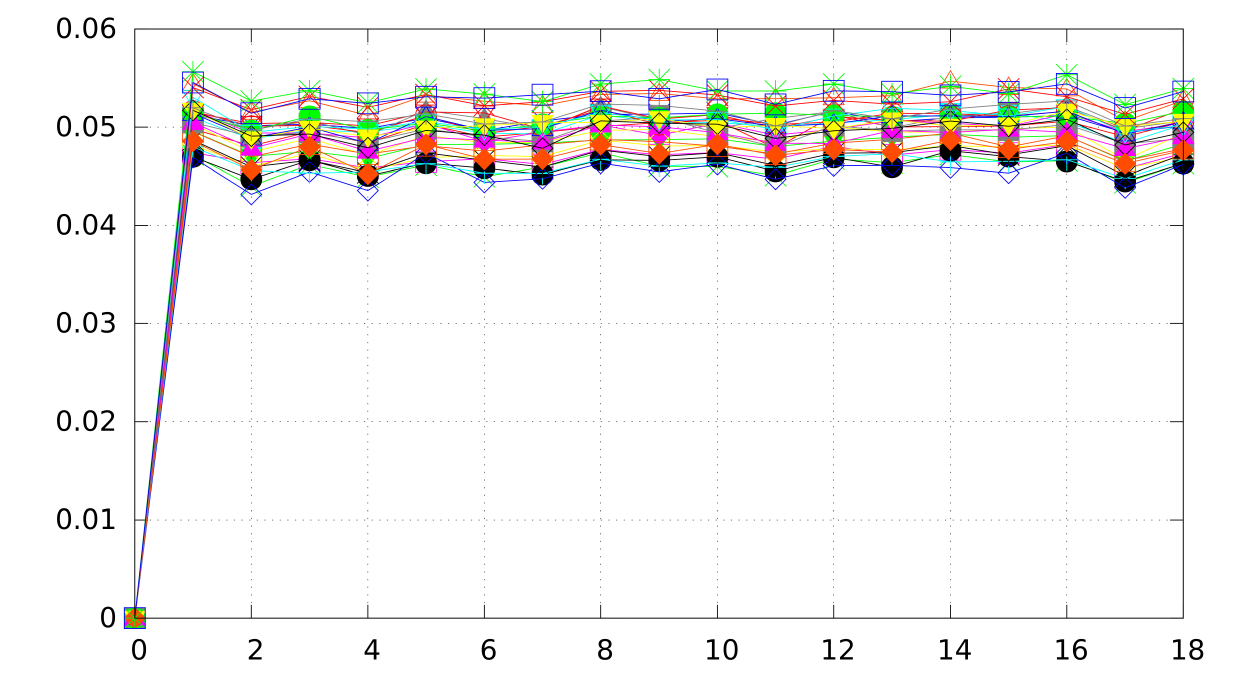}
	\ifpublic
	\caption[Within-class Hamming distance for antilles2, device 0, MPs 0--29.]{Within-class Hamming distance for antilles2, device 0, MPs 0--29.}
	\fi
	\label{fig:gpu-within-class-hd}
\end{figure}

\begin{figure}[h]
	\center
	\includegraphics[width=0.85\textwidth]{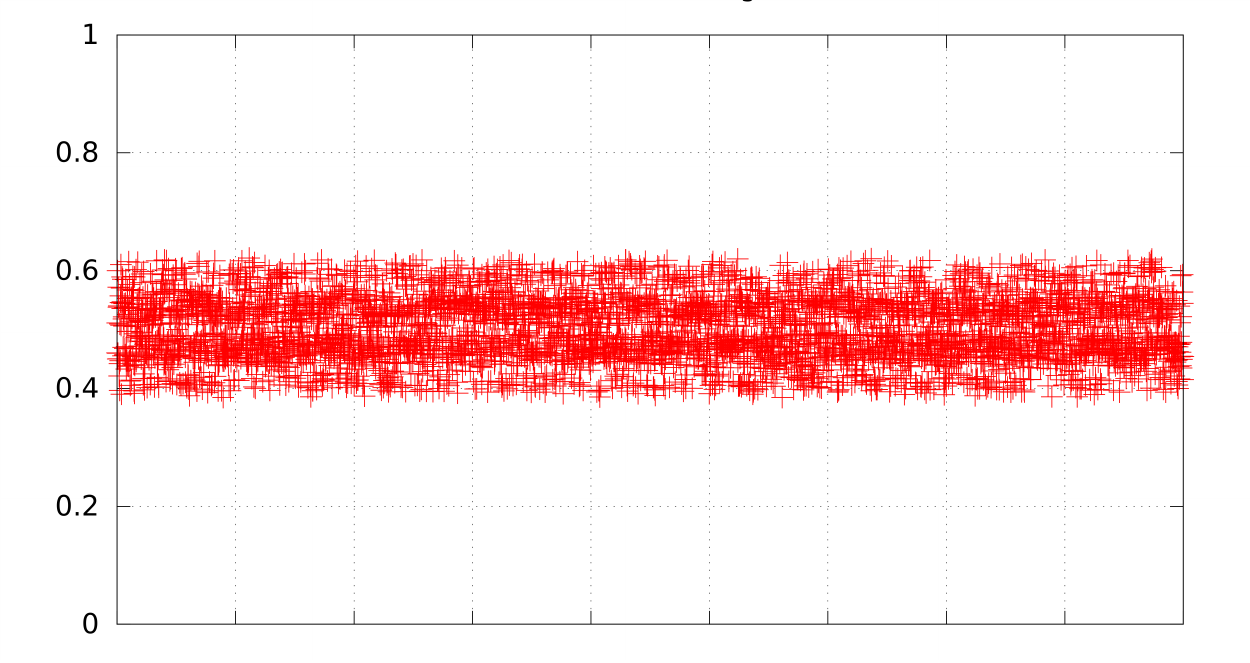}
	\caption[Between-class Hamming distance for all devices.]{Between-class Hamming distance for all devices.}
	\label{fig:gpu-between-class-hd}
\end{figure}

Figure~\ref{fig:gpu-within-class-hd} shows the within-class Hamming distance
from 18 different traces taken from each MP of device~0 on
\ifpublic
the machine ``antilles2''.
\fi
Each measurent is compared to the ``enrollment'' measurement~0.
The Hamming distance for each comparison is around 5\% which indicates that
the device can be identified with high accuracy.
Figure~\ref{fig:gpu-between-class-hd} shows the between-class Hamming distance
pairwise between all of our measurements.
The Hamming distance varied between 40\% and 60\%, which again indicates that
the difference between distinct devices is high and that each individual
device can be recognized accurately.
In particular, there is no general bias that maps certain bits of the SRAM to
the same value for all devices.
These
measurements and analysis show no obstacle to building a usable
PUF on top of these devices.

\FloatBarrier


\section{Discussion}
\label{sec::discussion}

Although we did not find a way to access and read either CPU registers or
CPU caches
before they are initialized, technically it would be possible to use them as
SRAM PUFs.
Thus, CPU vendors could enable these hardware features for the use as PUFs
probably with relatively small modifications to their chip designs.

As we explained, the situation seems to be different with at least
older-generation GPUs, yielding a usable PUF on the Nvidia GTX 295.

However, these SRAM PUFs in both CPU and GPU, if available to be read by
software either within the BIOS code or in the bootloader or operating system,
would not be protected against an attacker with any kind of root
access to the machine.
In case the attacker is able to read the PUF, he would be able to reproduce
the fingerprint and to impersonate the machine.
In case the attacker is able to deploy malware in the early boot process, he
would be able to manipulate the PUF state and thus he could influence, e.g.,
random number generation based on the PUF.
Strong software security is thus a prerequisite for truly secure use of these PUFs.

Our explorations on the GPU encountered challenges
when we upgraded to a different version of the Nvidia GPU drivers.
These drivers appeared to clear large amounts of GPU SRAM,
presumably in an effort to reduce the amount of undocumented behavior
exposed to GPU applications. Explicit memory zeroing is among the
recommended countermeasures against data leakage in~\cite{DLV13}.
Unfortunately, this also prevents using it as a PUF.
Furthermore, when we ran the same tests on a newer generation Nvidia card,
we were no longer able to retrieve the SRAM data. On ATI cards, we were
never able to read uninitialized SRAM data. This suggests that here, vendors are
actually trying to suppress this PUF-like behavior in their devices.

If CPU and GPU vendors decide to provide access to uninitialized SRAM state
for use as PUFs, further protection of their data is required.
However, data leakage should be prevented, as explained in~\cite{DLV13},
so maybe direct access is not the best solution.
An instruction-set extension as proposed in
\cite{OGMNPV13}, where the PUF data never leaves the CPU, could also be
applied to GPUs and seems to be the best way to implement this.

We have shown that the embedded SRAM in AMD64 CPUs, at least for the model we
tested, is indeed not usable as a PUF. For this, we have made modifications to
several open-source software packages. We release these modifications into the
public domain; they are available online.
We have also shown that PUFs are present in the Nvidia GTX 295 graphics card,
and conclude that they may be present in other graphics devices.

\subsection{Future work}

We have noticed the following phenomenon on a Lenovo ThinkPad X1 Carbon laptop,
2014 edition, with an Intel Core i7-4600U CPU and a 2560$\times$1440 screen;
note that this CPU contains a capable GPU embedded inside the CPU.
After the BIOS boot stage,
approximately the lower third of the screen is temporarily filled with
what appear to be randomly colored pixels.
This indicates possible presence of a PUF inside the video buffer on the GPU.
The obvious next step is to
use high-resolution photographic equipment to check
the Hamming distance between the colors after multiple power cycles.

\begingroup
\setlength{\emergencystretch}{8em}
\printbibliography
\endgroup

\clearpage


\vfill

\newpage

\appendix

\section{Software patches}
\label{app:src}

The kernel patch is shown in Listing~\ref{src::kernel}, the kernel module in
Listing~\ref{src::module}. The GRUB patch is shown in Listing~\ref{src::grub}.
The coreboot patch to output the registers is in
Listing~\ref{src::coreboot-registers}, and the coreboot patch to output the
stack space in cache is in Listing~\ref{src::coreboot-cache}.

\begin{listing}[h]
\begin{lstlisting}[
  basicstyle=\scriptsize\ttfamily,
  basewidth=0.5em,
  breaklines=true,
  numbers=left,
  stepnumber=1,
]
diff --git a/arch/x86/include/asm/i387.h b/arch/x86/include/asm/i387.h
index a850b4d..13cd1d1 100644
--- a/arch/x86/include/asm/i387.h
+++ b/arch/x86/include/asm/i387.h
@@ -15,6 +15,8 @@
 #include <linux/sched.h>
 #include <linux/hardirq.h>
 
+DECLARE_PER_CPU(struct i387_fxsave_struct, puf_data);
+
 struct pt_regs;
 struct user_i387_struct;

diff --git a/arch/x86/kernel/i387.c b/arch/x86/kernel/i387.c
index fa4ea09..48e1c00 100644
--- a/arch/x86/kernel/i387.c
+++ b/arch/x86/kernel/i387.c
@@ -157,6 +157,9 @@ static void __cpuinit init_thread_xstate(void)
 		xstate_size = sizeof(struct i387_fsave_struct);
 }
 
+DEFINE_PER_CPU(struct i387_fxsave_struct, puf_data);
+EXPORT_PER_CPU_SYMBOL(puf_data);
+
 /*
  * Called at bootup to set up the initial FPU state that is later cloned
  * into all processes.
@@ -188,6 +191,11 @@ void __cpuinit fpu_init(void)
 		cr0 |= X86_CR0_EM;
 	write_cr0(cr0);
 
+   asm volatile("FXSAVE %0\n" : "=m"
+         (per_cpu(puf_data, smp_processor_id())));  // save fpu registers
+
+   asm volatile("clts;");  // make sure to leave no trace
+
 	/*
 	 * init_thread_xstate is only called once to avoid overriding
\end{lstlisting}
\caption{The kernel patch for Linux Kernel version 3.15.7 to store the XMM registers.}
\label{src::kernel}
\end{listing}

\vfill

\begin{listing*}[h]
\begin{lstlisting}[
  basicstyle=\scriptsize\ttfamily,
  basewidth=0.5em,
  breaklines=true,
  numbers=left,
  stepnumber=1,
]
#include <linux/module.h>
#include <linux/proc_fs.h>
#include <linux/seq_file.h>
#include <asm/i387.h>

static struct proc_dir_entry* pufdata_file;

static int pufdata_show(struct seq_file *m, void *v)
{
  int i, cnt;

  for_each_cpu(i, cpu_present_mask) {
     seq_printf(m, "CPU %i:\n", i);

     for (cnt = 0; cnt < 64; cnt += 4)
       seq_printf(m, "%08x %08x %08x %08x\n", 
               per_cpu(puf_data, i).xmm_space[cnt + 3],
               per_cpu(puf_data, i).xmm_space[cnt + 2],
               per_cpu(puf_data, i).xmm_space[cnt + 1],
               per_cpu(puf_data, i).xmm_space[cnt + 0]);
     
     seq_printf(m, "\n");
  }

  return 0;
}

static int pufdata_open(struct inode *inode, struct file *file)
{
  return single_open(file, pufdata_show, NULL);
}

static const struct file_operations pufdata_fops = {
  .owner  = THIS_MODULE,
  .open  = pufdata_open,
  .read = seq_read,
  .llseek  = seq_lseek,
  .release  = single_release,
};

static int __init pufdata_init(void)
{
  pufdata_file = proc_create("pufdata", 0, NULL, &pufdata_fops);

  if (!pufdata_file) {
    return -ENOMEM;
  }

  return 0;
}

static void __exit pufdata_exit(void)
{
  remove_proc_entry("pufdata", NULL);
}

module_init(pufdata_init);
module_exit(pufdata_exit);

\end{lstlisting}
\caption{The kernel module for Linux Kernel version 3.15.7 to access the stored register values.}
\label{src::module}
\end{listing*}

\begin{listing*}[h]
\begin{lstlisting}[
  basicstyle=\scriptsize\ttfamily,
  basewidth=0.5em,
  breaklines=true,
  numbers=left,
  stepnumber=1,
]
diff --git a/grub-core/kern/main.c b/grub-core/kern/main.c
index 9cad0c4..0cbd8f0 100644
--- a/grub-core/kern/main.c
+++ b/grub-core/kern/main.c
@@ -29,6 +29,7 @@
 #include <grub/command.h>
 #include <grub/reader.h>
 #include <grub/parser.h>
+#include <grub/time.h>
 
 #ifdef GRUB_MACHINE_PCBIOS
 #include <grub/machine/memory.h>
@@ -269,11 +270,47 @@ grub_main (void)
 
   grub_boot_time ("After machine init.");
 
+  unsigned char puf_result[128];
+  int puf_i;
+  for (puf_i = 0; puf_i < 128; ++puf_i) {
+    puf_result[puf_i] = 0xca;
+  }
+  asm(
+       "mov %%cr0,   %%eax;"
+       "and $0xFFFB, %%ax;" /* Clear coprocessor emulation CR0.EM */
+       "or  $0x2,    %%ax;" /* Set coprocessor monitoring CR0.MP */
+       "mov %%eax,   %%cr0;"
+       "mov %%cr4,   %%eax;"
+       "or $0x0600,  %%ax;" /* Set CR4.OSFXSR and CR4.OSXMMEXCPT at the same time */
+       "mov %%eax,   %%cr4;"
+       :::"%eax","%ax"
+     );
+  asm(
+       "movupd %%xmm0, 0(%0);"
+       "movupd %%xmm1, 16(%0);"
+       "movupd %%xmm2, 32(%0);"
+       "movupd %%xmm3, 48(%0);"
+       "movupd %%xmm4, 64(%0);"
+       "movupd %%xmm5, 80(%0);"
+       "movupd %%xmm6, 96(%0);"
+       "movupd %%xmm7, 112(%0);"
+       ::"r" (puf_result)
+     );
+  for (puf_i = 0; puf_i < 128; ++puf_i) {
+    grub_printf ("%02x ", puf_result[puf_i]);
+    if (puf_i % 16 == 15) {
+      grub_printf ("\n");
+    }
+  }
+  grub_printf ("\n\n");
+
   /* Hello.  */
   grub_setcolorstate (GRUB_TERM_COLOR_HIGHLIGHT);
   grub_printf ("Welcome to GRUB!\n\n");
   grub_setcolorstate (GRUB_TERM_COLOR_STANDARD);
 
+  grub_printf("Will now sleep for 60 seconds\n\n");
+  grub_millisleep(60000);
   grub_load_config ();
 
   grub_boot_time ("Before loading embedded modules.");
\end{lstlisting}
\caption{The GRUB patch for GRUB version 2.02-beta2 to output the XMM registers.}
\label{src::grub}
\end{listing*}

\begin{listing*}[h]
\begin{lstlisting}[
  basicstyle=\scriptsize\ttfamily,
  basewidth=0.5em,
  breaklines=true,
  numbers=left,
  stepnumber=1,
]
diff --git a/src/mainboard/asrock/e350m1/romstage.c b/src/mainboard/asrock/e350m1/romstage.c
index 2913c08..89129cf 100644
--- a/src/mainboard/asrock/e350m1/romstage.c
+++ b/src/mainboard/asrock/e350m1/romstage.c
@@ -45,6 +45,23 @@ void cache_as_ram_main(unsigned long bist, unsigned long cpu_init_detectedx)
 {
 	u32 val;
 
+	u32 puf_i;
+	unsigned char puf_result[128];
+	for (puf_i = 0; puf_i < 128; ++puf_i) {
+		puf_result[puf_i] = 0xca;
+	}
+	asm(
+		"movupd %%xmm0, 0(%0);"
+		"movupd %%xmm1, 16(%0);"
+		"movupd %%xmm2, 32(%0);"
+		"movupd %%xmm3, 48(%0);"
+		"movupd %%xmm4, 64(%0);"
+		"movupd %%xmm5, 80(%0);"
+		"movupd %%xmm6, 96(%0);"
+		"movupd %%xmm7, 112(%0);"
+		::"r" (puf_result)
+	);
+
 	/*
 	 * All cores: allow caching of flash chip code and data
 	 * (there are no cache-as-ram reliability concerns with family 14h)
@@ -74,6 +91,14 @@ void cache_as_ram_main(unsigned long bist, unsigned long cpu_init_detectedx)
 	printk(BIOS_DEBUG, "cpu_init_detectedx = %08lx \n", cpu_init_detectedx);
 
 	post_code(0x35);
+
+	for (puf_i = 0; puf_i < 128; ++puf_i) {
+		printk(BIOS_DEBUG, "%02x ", puf_result[puf_i]);
+		if (puf_i % 16 == 15) {
+			printk(BIOS_DEBUG, "\n");
+		}
+	}
+
 	printk(BIOS_DEBUG, "agesawrapper_amdinitmmio ");
 	val = agesawrapper_amdinitmmio();
 	if (val)
\end{lstlisting}
\caption{The coreboot patch to output the XMM registers.
	The patch is based on coreboot git commit c86762657dc7013a56b1d281286789dae17ad936.}
\label{src::coreboot-registers}
\end{listing*}

\begin{listing*}[h]
\begin{lstlisting}[
  basicstyle=\scriptsize\ttfamily,
  basewidth=0.5em,
  breaklines=true,
  numbers=left,
  stepnumber=1,
]
diff --git a/src/mainboard/asrock/e350m1/romstage.c b/src/mainboard/asrock/e350m1/romstage.c
index 2913c08..5e02764 100644
--- a/src/mainboard/asrock/e350m1/romstage.c
+++ b/src/mainboard/asrock/e350m1/romstage.c
@@ -44,6 +44,7 @@
 void cache_as_ram_main(unsigned long bist, unsigned long cpu_init_detectedx)
 {
 	u32 val;
+	u32 puf_i;
 
 	/*
 	 * All cores: allow caching of flash chip code and data
@@ -74,6 +75,13 @@ void cache_as_ram_main(unsigned long bist, unsigned long cpu_init_detectedx)
 	printk(BIOS_DEBUG, "cpu_init_detectedx = %08lx \n", cpu_init_detectedx);
 
 	post_code(0x35);
+
+	printk(BIOS_DEBUG, "%p\n", &puf_i);
+
+	for (puf_i = 0; puf_i < 16384; ++puf_i) {
+		printk(BIOS_DEBUG, "%08x\n", *((&puf_i) - puf_i));
+	}
+
 	printk(BIOS_DEBUG, "agesawrapper_amdinitmmio ");
 	val = agesawrapper_amdinitmmio();
 	if (val)
diff --git a/src/vendorcode/amd/agesa/f14/gcccar.inc b/src/vendorcode/amd/agesa/f14/gcccar.inc
index 40e0e31..5bd45f9 100644
--- a/src/vendorcode/amd/agesa/f14/gcccar.inc
+++ b/src/vendorcode/amd/agesa/f14/gcccar.inc
@@ -1542,11 +1542,11 @@ ClearTheStack:                          # Stack base is in SS, stack pointer is
     jne 1f
         cld
         mov     %edi, %esi
-        rep     lodsl (%esi)    # Pre-load the range
+        #rep     lodsl (%esi)    # Pre-load the range
         xor     %eax, %eax
         mov     %bx, %cx
         mov     %edi, %esi                # Preserve base for push on stack
-        rep     stosl (%edi)    # Clear the range
+        #rep     stosl (%edi)    # Clear the range
         movl     $0x0ABCDDCBA, (%esp) # Put marker in top stack dword
         shl     $2, %ebx                  # Put stack size and base
         push    %ebx                     #  in top of stack
\end{lstlisting}
\caption{The coreboot patch to output the cache-as-RAM stack space.
	The patch is based on coreboot git commit c86762657dc7013a56b1d281286789dae17ad936.}
\label{src::coreboot-cache}
\end{listing*}

\end{document}